\documentclass[12pt]{article}
\usepackage[dvips]{graphics}

\begin{document}

\begin{titlepage}
\begin{center}

\hfill    FTPI-MINN-06/21\\
\hfill     hep-ph/0606158\\

\quad\\
\quad\\

\vskip 1cm

{\large \bf CKM pattern from localized generations in extra dimension}

\vskip 1cm

C.~Matti\footnote[2]{cyril.matti@a3.epfl.ch}\footnote[5]{On leave from Ecole Polytechnique F\'ed\'erale de Lausanne, 1015 Lausanne, Switzerland}

\vskip 0.05in

{\em William I. Fine Theoretical Physics Institute, University of Minnesota, Minneapolis, MN 55455, USA}

\end{center}

\vskip 3cm

\begin{abstract}
We revisit the issue of the quark masses and mixing angles in the framework of large extra dimension. We consider three identical standard model families resulting from higher-dimensional fields localized on different branes embedded in a large extra dimension. Furthermore we use a decaying profile in the bulk different form previous works. With the Higgs field also localized on a different brane, the hierarchy of masses between the families results from their different positions in the extra space. When the left-handed doublet and the right-handed singlets are localized with different couplings on the branes, we found a set of brane locations in one extra dimension which leads to the correct quark masses and mixing angles with the sufficient strength of \textit{CP}-violation. We see that the decaying profile of the Higgs field plays a crucial role for producing the hierarchies in a rather natural way.
\end{abstract}

\end{titlepage}


\section{Introduction}
\label{intro} 

The origin of the quark masses and their mixing angles is still one of the mysteries of the standard model (SM). The great hierarchy of scale between the SM Yukawa couplings should be justified in a fundamental theory. We usually thought about low-energy constants as resulting from some breaking of higher symmetry operators. Several years ago, new mechanisms were proposed by \cite{Antoniadis, Hamed-Schmaltz, Dvali-Shifman} to explain the hierarchy of the Yukawa couplings from geometry of extra space without any new flavor symmetries.

In these models, we consider the SM four-dimensional fermionic fields as zero modes of higher-dimensional fields localized at different positions in an extra dimension. The hierarchies come from the overlap of the extra-dimensional part of the wave functions, which is exponentially small according to their respective distances in the extra dimension. In this framework we can easily understand the origin of the well-known nearest neighbor mixing: fields which are literally closer in the extra space are coupled stronger than distant fields. This can explain, at least at the conceptual level, the pattern of the SM quark masses as well as their mixing.

Since the work of \cite{Mirabelli-Schmaltz,tait} we know that under certain assumptions on the profile functions we can find realistic parameters of localization in the background of \cite{Hamed-Schmaltz} which reproduce the correct quark masses and the correct magnitude of the Cabibbo-Kobayashi-Maskawa (CKM) matrix elements with one extra dimension. From \cite{Branco-Rebelo}, we know that with two extra dimensions we can also reproduce the correct amount of \textit{CP}-violation from the complex phase of the CKM matrix. A careful analysis of the neutrino masses and the lepton flavor violation were also done in \cite{lepton}.

These authors used Gaussian functions for the wave functions overlap, an assumption which cannot be applied for the case where the distances between localized branes are larger than the brane widths. This is the case when the fields are localized on separate branes and not in one `fat' brane like in \cite{Hamed-Schmaltz}. The zero mode is effectively exponential in the bulk, and thus the Gaussian functions are not relevant. We will then follow \cite{Dvali-Shifman} and consider the wave functions overlap as being due to their exponential tails. By this way, we use a decaying profile of the zero modes different from those used in previous papers \cite{Hamed-Schmaltz,Mirabelli-Schmaltz,Branco-Rebelo,lepton}.

\begin{figure}[tbp]
\centerline{
\resizebox{1\textwidth}{!}{\includegraphics{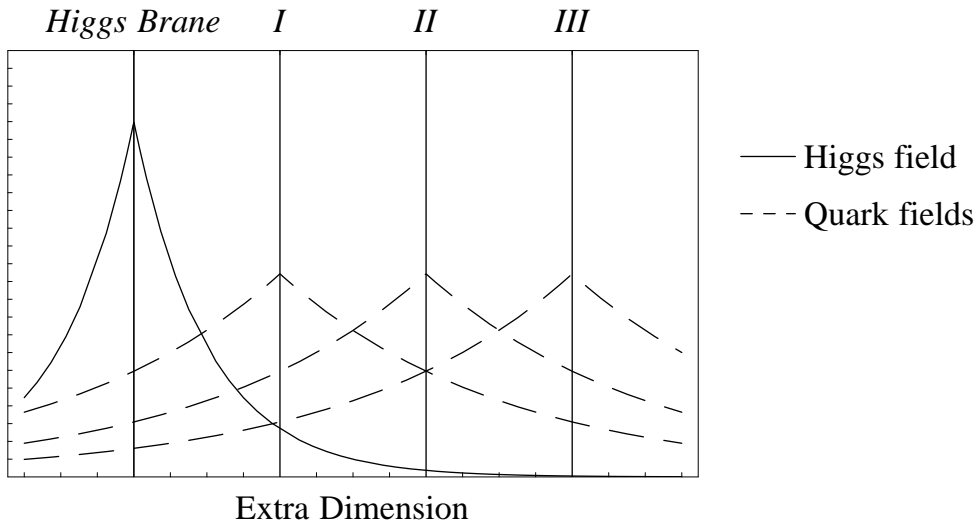}}
}
\caption{Extra-dimensional part of arbitrary zero modes (with exponential profiles) for the localized Higgs field and three identical generations of quark fields localized on different branes (I, II and III). The distances between the branes produce an exponentially small overlap of the wave functions. \textit{O}(1) parameters can thus provide a large hierarchy of scale}
\label{graph1}
\end{figure}

Another distinct feature of \cite{Dvali-Shifman}, which plays a more interesting role, comes from the consideration of the electroweak (EW) symmetry breaking localized on a separate brane. This allows us to consider three identical generations of quarks localized on three different branes. The hierarchy of masses between the different generations is then produced by the decaying Higgs profile, while their mixing contain also suppression from the distances between the generations; see Fig.~\ref{graph1}. With this pattern, the texture of the mass matrix in the weak basis is more natural than the one obtained in \cite{Mirabelli-Schmaltz,Branco-Rebelo}.

Now the question remains whether a realistic set of locations can be found to reproduce the correct experimental masses and mixing angles. Our framework provides a non-trivial relationship between the masses and the mixing. In this paper, we address the question whether this relationship is compatible with the SM quark masses and the CKM matrix.

We found that it is possible to match correctly the experimental data of the quark sector with three identical generations localized in one extra dimension. For this we need to consider the SM left-handed doublet and the right-handed singlets coupled with different strengths to the brane. With all the fields localized on the same side of the EW symmetry breaking brane, we found a set of dispersion coefficients and brane distances in the extra dimension reproducing the quark masses and the CKM matrix with the correct strength of \textit{CP}-violation.

The paper is organized as follows. In Sect.~\ref{dsscenario} we sum up the main points of \cite{Dvali-Shifman}. In Sect.~\ref{parameterspace} we present the constraints on the parameters of such a mechanism. In Sect.~\ref{realisticparameters} we briefly recall the SM parameters and present results of our effort to reproduce these values. We then conclude in Sect.~\ref{conclusion}.


\section{DS scenario}
\label{dsscenario}

Let us summarize the main points of the Dvali-Shifman (DS) scenario \cite{Dvali-Shifman}. We consider our universe as $(4+n)$-dimensional, with topology $M_4 \times S$, the usual Minkowski space times an extra space $S$ of dimension $n$ with size $L$. The observed SM four-dimensional fields are zero modes of higher-dimensional fields localized on different branes. A complete theory is given by some field theory and geometry of multiple brane state embedded in $S$, which can be a compactified dimension or a `fatter' brane. The fields building the branes are distinct from the matter fields. Each brane (which could be D-branes or topological defects for instance) is responsible for localizing one SM generation. In order to allow interactions among the fermionic fields, the gauge fields are free to propagate throughout $S$. This constrains $L$ to be $\sim 1/10 \ {\rm TeV}^{-1}$ to avoid flavor exchange violation at tree level. Depending on the underlying model considered, this constraint could even require $L$ to be much smaller \cite{Delgado}. We will treat it as a free parameter depending on the complete theory. A mechanism of embedding multiple brane states in a compact space were proposed e.g. in \cite{Dvali-Shifman}.

We will now consider the case of one extra dimension $(n=1)$. One brane inside the extra dimension is responsible for trapping fermionic fields with all the quantum numbers of one SM generation. Thus, a three brane state gives rise to the three known SM generations. One crucial point of the DS scenario is to consider the EW symmetry breaking happening on a separate brane. A possible mechanism to localize the Higgs field on a brane is proposed in \cite{Dvali-Shifman}. In general, producing such condensate results in an exponential shape of the Higgs profile in the bulk. Far from the source, the Higgs will be seen as ${\rm e}^{-\overline{m}_h|y|}$, where $\overline{m}_h$ is a coefficient of dispersion setting the thickness of the zero mode and $y$ is the extra dimension. The three generations will then obtain an exponential hierarchy of masses if they are localized at different positions in the extra dimension. This reproduces the well-known hierarchy of masses of the SM generations. The masses are suppressed according to the overlap of the square of the five-dimensional part of the fermionic zero mode with the decaying Higgs profile. The mixing across the generations contains the additional suppression of the fermionic zero mode overlap. The further the generations are from each other, the more suppressed the mixing will be; see Fig.~\ref{graph1}. This is in agreement with the experimental pattern.

We will have the following features independently of the underlying higher-dimensional theory. Let us start from the five-dimensional fermionic fields $F$ and $F^{\rm c}$, where the upperscript c stands for the charge conjugate. Only one chirality can be trapped on the brane and so the SM right-handed singlets are provided by $F_i^{\rm c}=U_i^{\rm c}$ and $F_i^{\rm c}=D_i^{\rm c}$, while the \textit{SU}(2) left-handed doublets are produced by $F_i=Q_i$. The Yukawa terms are given by
\begin{equation}
S_{\rm Yuk}=\int {\rm d}^5 x gH\overline{F_1}F_2^{\rm c},
\end{equation}
where $g$ is the five-dimensional Yukawa coupling and $H$ the five-dimensional Higgs field. As the fermionic fields are localized on a brane, the following expansion is valid:
\begin{eqnarray}
S_{\rm Yuk} &=& \int {\rm d}^5x \ gH\overline{F_1}F_2^{\rm c} \nonumber\\
&=& g\int {\rm d}y \ \Omega_1(y-y_1) \Omega_{2}(y-y_2) {\rm e}^{-\overline{m}_h|y|} \nonumber\\
&& \times\int {\rm d}^4x \ hf_1f_2+\cdots,
\end{eqnarray}
where $y$ is the extra dimension, $f_{1,2}$ are zero modes of the four-dimensional Dirac operator and $h$ is the four-dimensional part of the zero mode of the localized Higgs field. The $\Omega_{1,2}$ functions are the extra-dimensional part of the five-dimensional fermionic fields localized at $y_{1,2}$.
When considering only the zero modes and generalizing to several generations,
\begin{equation}
S_{\rm Yuk} = \sum_{ij}\lambda_{ij} \int {\rm d}^4x \ hf_if_j,
\end{equation}
where $i,j=1,2,3$ for three generations. Thus, the effective four-dimensional Yukawa terms are
\begin{equation}
\lambda_{ij}=g\int {\rm d}y \ \Omega_i(y-y_i) \Omega_{j}(y-y_j){\rm e}^{-\overline{m}_h|y|} .
\label{yukawa}
\end{equation}
The $\Omega$ functions are given by the solution of the five-dimensional part of the Dirac operator. When considering only the left-handed chirality:
\begin{equation}
(\partial+\varphi(s))\Omega=0,
\label{equationomega}
\end{equation}
where $\varphi(s)$ is a five-dimensional scalar field with a domain wall profile responsible for trapping the fermionic fields (see \cite{Hamed-Schmaltz} for details). Equation~(\ref{equationomega}) gives
\begin{equation}
\Omega(y)\sim {\rm e}^{-\int_0^y\varphi(s){\rm d}s}.
\end{equation}
Thus, in the bulk and localizing the field at $y_i$,
\begin{equation}
\Omega(y)\sim {\rm e}^{-\overline{m}|y-y_i|},
\end{equation}
with $\overline{m}$ fitting the dispersion of the zero mode. Setting the Higgs field to its vacuum expectation value $\langle h\rangle$, we have the following mass matrices $M_{(u,d)}$ for the up- and down-type quarks, respectively:
\begin{eqnarray}
\lefteqn{M_{ij(u,d)}=\langle h\rangle \lambda_{ij}}\nonumber\\
&&=\rho \int {\rm d}y \ {\rm exp}\left\lbrace -\overline{m}_h|y|-\overline{m}_q|y-y_i|-\overline{m}_{(u,d)}|y-y_j|\right\rbrace, \quad
\label{mass matrix}
\end{eqnarray}
where $\rho$ contains the five-dimensional Yukawa coupling, the Higgs vacuum expectation value and some normalization parameters of the $\Omega$ functions. The $y_i$ are the locations in the extra dimension of the branes. We made the particular choice of localizing the Higgs field at $y_h=0$, but this plays no role since an overall translation of all the locations lets the integral in (\ref{mass matrix}) invariant. The mass terms $\overline{m}_{(h,q,u,d)}$ represent the width of the respective zero mode profiles which depends on the process of localization. As in the general case the Yukawa couplings have complex entries, we consider the elements of (\ref{mass matrix}) to have arbitrary complex phases.

We see that $M_{(u,d)}$ is invariant under
\begin{equation}
\overline{m}\rightarrow  \overline{m}\mu,\ \ \ \ y_i\rightarrow y_i \mu^{-1} \ \ \ \ {\rm and}\ \ \ \ \rho \rightarrow \rho\mu,
\label{scale}
\end{equation}
and, thus, the scale $\mu$ of the different parameters can be chosen arbitrarily according to the requirement of the full theory.


\section{Parameter space}
\label{parameterspace}

We are interested whether the mass matrix (\ref{mass matrix}) can reproduce the quark masses and mixing angles. Due to the flavor symmetry, we can redefine the fields to bi-diagonalize the matrix, the diagonal elements thus giving the quark masses. The CKM matrix elements come from the mismatch between the weak and the mass basis. An easy way of computing such elements is to consider the following hermitian matrices:
\begin{eqnarray}
H_u&\equiv& M_uM_u^{\dagger}, \nonumber\\ H_d&\equiv& M_dM_d^{\dagger}.
\end{eqnarray}
Their eigenvalues give the square of the quark masses. The product of the eigenvectors of $H_u$ with the eigenvectors of $H_d$ gives the CKM matrix. The global $U(1)$ symmetry of the fermionic fields allows us to perform the following transformation:
\begin{eqnarray}
H_u&\rightarrow& K^{\dagger}H_uK, \nonumber\\ H_d&\rightarrow& K^{\dagger}H_dK,
\end{eqnarray}
with $K$ a diagonal complex matrix of pure phases. Thus, we can suppress all arbitrary complex phases of $H_u$ coming from (\ref{mass matrix}), while $H_d$ remains with arbitrary phases. We can then find the eigenvalues by the orthogonal transformations
\begin{eqnarray}
O_u^{\dagger}~H_u~O_u={\rm diag}(m_u^2,m_c^2,m_t^2),\nonumber\\
O_d^{\dagger}K'^{\dagger}~H_d~K'O_d={\rm diag}(m_d^2,m_s^2,m_b^2),
\end{eqnarray}
where $K'$ is an arbitrary diagonal complex matrix of pure phases which suppresses the arbitrary phases of $H_d$.
The quark mixing are given by the CKM matrix defined by
\begin{equation}
V_{\rm CKM}=O_u^{\dagger}K'O_d.
\label{CKM}
\end{equation}
Since an overall redefinition of the phases has no influence on the magnitude of the CKM matrix elements, we can choose $K'={\rm diag}(1,{\rm e}^{i\phi},{\rm e}^{i\sigma})$ without any loss of generality. Thus, we can represent all the arbitrary weak phases of (\ref{mass matrix}) with only two parameters, $\phi$ and $\sigma$.

In the work of \cite{Mirabelli-Schmaltz} and \cite{Branco-Rebelo}, the choice of the weak basis plays a special role. Their specific background does not allow for any natural basis and they must fit the SM parameters by some specific configuration of the mass matrix. In our pattern, the localization of the Higgs field brings about a more natural texture. We can effectively choose a weak basis where the diagonal elements are decreasing according to the Higgs profile, and where the off-diagonal elements are also suppressed by the fermionic overlap. This leads to the following texture:
\begin{eqnarray}
\left( \begin{array}{ccc} A & a & b \\
\tilde{a} & B & c \\ \tilde{b} & \tilde{c} & C \end {array} \right),
\label{texture}
\end{eqnarray}
where, in general, $A>B>C$, $a>b>c$ and $\tilde{a}>\tilde{b}>\tilde{c}$ in magnitude.
 
However, the off-diagonal elements cannot be much smaller than the diagonal elements. Effectively, as most of the down quarks are lighter than the up quarks, we need to have $\overline{m}_d > \overline{m}_u$ to obtain this difference of masses as well as the CKM matrix. (Another possibility to render the down quarks lighter would be to have $g$ (the Yukawa coupling of (\ref{yukawa})) smaller for the down quarks. However, we see from (\ref{CKM}) that to have a non-trivial CKM matrix, the matrix elements of $M_u$ and $M_d$ cannot be proportional.) The difference $\overline{m}_d > \overline{m}_u$ is allowed by the fact that the up-type and the down-type singlets are generated by different five-dimensional fields which can be coupled to the brane in different ways. With this mechanism, the elements of $M_d$ are in general smaller than the elements of $M_u$. Therefore, the special feature of the SM quark masses to have $m_u<m_d$ while we have $m_c>m_s$ and $m_t>m_b$ requires for our mechanism to obtain this case with an adequate behavior of the off-diagonal elements of the mass matrices. This forces us to have the off-diagonal elements of the same magnitude as the diagonal elements.


\section{Realistic parameters}
\label{realisticparameters}

First of all, we need to run the quark masses to a common scale. We choose this scale to be $m_t$, but a further running to a higher scale can be contained in the arbitrary power of the five-dimensional Yukawa coupling (\ref{scale}) and thus plays no role. We used the running constants $\eta _i \equiv m_i(m_i)/m_i(m_t)$ for $i=c,\ b,\ t$ and $\eta _i \equiv m_i(\mbox{2 GeV)}/m_i(m_t)$ for $i=u,\ d,\ s$. The different values of $\eta$ were computed to one loop in QED and three loops in QCD \cite{Mirabelli-Schmaltz,Branco-Rebelo,eta1,eta2} and are given by
\begin{eqnarray}
\eta_u=1.84, \ & \eta_d=1.84, \ & \eta_s=1.84, \nonumber \\
\eta_c=2.17, \ & \eta_b=1.55, \ & \eta_t=1.00.
\label{eta}
\end{eqnarray}
For the allowed SM experimental values, we used \cite{pdg}
\begin{eqnarray}
m_u&=&1.5 \; \mbox{to} \; 4 \; \mbox{MeV}, \nonumber \\
m_c&=&1150 \; \mbox{to} \; 1350 \; \mbox{MeV}, \nonumber \\
m_t&=&166 000 \pm 5 000 \; \mbox{MeV}, \nonumber \\
m_d&=&4 \; \mbox{to} \; 8 \; \mbox{MeV}, \nonumber \\
m_s&=&80  \; \mbox{to} \; 130 \; \mbox{MeV}, \nonumber \\
m_b&=&4100 \; \mbox{to} \; 4400 \; \mbox{MeV},
\label{SMvalues}
\end{eqnarray}
where the $u$-, $d$- and $s$-quark masses are the current quark masses estimated in the $\overline{\rm MS}$ scheme at a scale $\mu \approx 2$ GeV and the $c$-, $b$-, $t$-quark masses are the running masses in the $\overline{\rm MS}$ scheme evaluated at a scale equal to their $\overline{\rm MS}$ mass. The running of the CKM matrix elements is small comparing to their incertitude and, thus, we do not take it in account. From \cite{pdg}, we have
\begin{eqnarray}
|V_{us}| & = & 0.221  \ {\rm to} \  0.227, \nonumber \\ 
|V_{ub}| & = & 0.0029 \  {\rm to}  \ 0.0045, \nonumber \\ 
|V_{cb}| & = & 0.039 \  {\rm to}  \ 0.044 .
\end{eqnarray}

To find a set of parameters fitting these values, we proceeded to a numerical exploration of the parameter space. Starting from an arbitrary point we then moved in the direction where the parameters fit the SM values better. This method tells us nothing about the uniqueness and the degeneracy of the solutions, but it is adequate for the simple purpose of finding at least one correct set of parameters which proves compatibility of our scenario with the experimental data. We proceeded in this way after failing to find analytical solutions for mass matrices with textures as described in Sect.~\ref{parameterspace}.

We used (\ref{mass matrix}) for the mass matrix with the arbitrary value of $\rho = 247 000 \ \mu$, where $\mu$ is a dimensionful parameter. It means that all parameters are given in units of $\mu$, which can be chosen according to the theory considered. This could be probably $\sim$ TeV. One condition on this scale is to have the distances between the branes $|y_i-y_j|$ much bigger than the brane widths, in order to justify the approximation of the exponential function. (This particular choice of $\rho$, equal for the matrices $M_u$ and $M_d$, also means that we implicitly consider different five-dimensional Yukawa couplings for the up and the down fields.)

\begin{figure}[tbp]
\centerline{
\resizebox{1\textwidth}{!}{\includegraphics{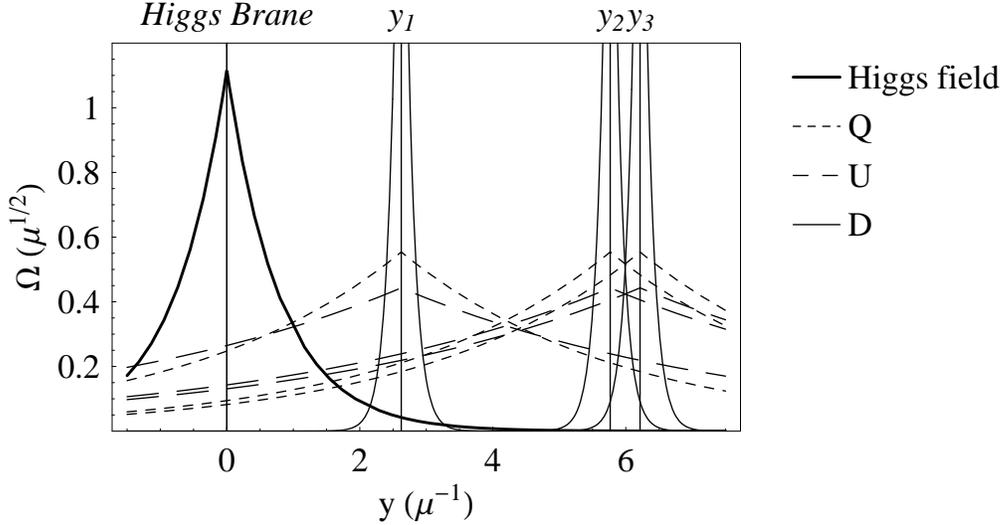}}
}
\caption{Decaying profile in the extra dimension provided by the set of parameters (\ref{results1}) and (\ref{results2}) for the zero modes of the Higgs field and three identical generations of quarks. $Q$ stands for the profile of the \textit{SU}(2) left-handed doublets, while $U$ and $D$ stand for the profile of the \textit{SU}(2) right-handed singlets, up-type and down-type respectively. The figure is drawn in units of $\mu$ and the functions are normalized (squared)}
\label{graph2}
\end{figure}

After a couple of attempts, we found for the three brane locations:
\begin{eqnarray}
y_1 &=& 2.6238 \; \mu^{-1},\nonumber \\
y_2 &=& 5.7661 \; \mu^{-1},\nonumber \\
y_3 &=& 6.2148 \; \mu^{-1},\nonumber \\
\label{results1}
\end{eqnarray}
where the Higgs brane is localized at $y=0$. For the parameters setting the widths of the zero modes we have
\begin{eqnarray}
\overline{m}_h &=& 1.2489 \; \mu,\nonumber \\
\overline{m}_q &=& 0.3070 \; \mu,\nonumber \\
\overline{m}_u &=& 0.1963 \; \mu,\nonumber \\
\overline{m}_d &=& 7.6006 \; \mu.
\label{results2}
\end{eqnarray}
This gives the mass matrices
\begin{eqnarray}
M_u &=&\left( \begin{array}{ccc} 11 8901 & 64983.9 & 59509.5 \\
46376.5 & 25789.4 & 23627.6 \\ 40418.7 & 22490.2 & 20 608 \end {array} \right) 
\; \mbox{MeV}, \nonumber \\
M_d &=&\left( \begin{array}{ccc} 2418.41 & 19.2758 & 9.58932 \\
949.553 & 47.7699 & 25.0693 \\ 827.35 & 42.8544 & 27.2746 \end {array} \right)
\; \mbox{MeV}.
\end{eqnarray}
When taking in account the running constants $\eta_i$ (\ref{eta}), we obtain the quark masses:
\begin{eqnarray}
m_u&=& 2.68 \; \mbox{MeV}, \nonumber \\
m_c&=& 1268 \; \mbox{MeV}, \nonumber \\
m_t&=& 166852 \; \mbox{MeV}, \nonumber \\
m_d&=& 5.62 \; \mbox{MeV}, \nonumber \\
m_s&=& 102 \; \mbox{MeV}, \nonumber \\
m_b&=& 4227 \; \mbox{MeV}.
\end{eqnarray}
This is in the range of values allowed by (\ref{SMvalues}). Our results (\ref{results1}) and (\ref{results2}) are illustrated in Fig.~\ref{graph2}. This solution is probably not unique. 

Furthermore, with the choice of weak phases 
\begin{equation}
\phi=0.5297 \ \ \ \ {\rm and}\ \ \ \ \sigma=0.5863,
\end{equation}
we find the following CKM matrix:
\begin{equation}
| V_{\rm CKM} | = \left( \begin{array}{ccc} 0.9745  & 0.2244 & 0.0036 \\
0.2243 & 0.9736 & 0.0419 \\ 0.0085 & 0.0412 & 0.9991 \end{array} \right),
\end{equation}
with the strength of \textit{CP}-violation
\begin{equation}
J\equiv |\mbox{Im} (V_{ub}V_{cs}V_{us}^*V_{cb}^*)|= 2.93 \times 10^{-5},
\end{equation}
which agrees with the experimental value \cite{pdg} $J=(2.88\pm0.33)\times 10^{-5}$. We see that we can produce a correct amount of \textit{CP}-violation with only one extra dimension, as opposed to the situation of \cite{Hamed-Schmaltz}, where only two extra dimensions were found to be adequate \cite{Branco-Rebelo}.

One last comment regarding the results (\ref{results1}) and (\ref{results2}). We have
\begin{equation}
\overline{m}_u \vert y_2-y_3\vert\sim0.09.
\end{equation}
One might then argue that the Gaussian parts of the wave functions inside the branes could play a role, since the brane width $\delta$ can usually be related with the dispersion coefficient such as $\overline{m}^{-1}\sim\delta$. However this influence should not be an obstacle since it will only give a little bit more suppression which can be compensated by putting the branes closer. A detailed analysis of the situation depends on the underlying higher-dimensional theory and, thus, must be done during the building of a concrete model. In any case, our study demonstrates compatibility of such models with the quark masses and mixing.


\section{Conclusion}
\label{conclusion}

With the set of parameters (\ref{results1}) and (\ref{results2}), we see that models with localized fields in extra-dimensional theories could be an explanation for the hierarchies of the SM Yukawa couplings. We see that the particular SM situation of $m_u<m_d$ is a non-trivial configuration for the process of producing hierarchy with exponentially small wave functions overlap but is nevertheless compatible with such process. It is encouraging to find a possible way of producing the correct hierarchy in a rather natural way and it should be a motivation to search a full realistic model.

The study of the lepton case represents a challenge for the future. The relatively large mixing of the lepton sector comparing to the neutrino masses, as well as the possible Majorana nature of these particles, must be studied in detail to see whether the DS approach to the hierarchy problem can provide a sufficient explanation.


\section*{Acknowledgements}

I would like to thank Mikhail Shifman for his wise advice and his helpful discussions. I also thank Mikhail Shaposhnikov for having made it possible to realize this work, Roberto Auzzi and Tony Gherghetta for helpful discussions and the FTPI staff at the University of Minnesota for hospitality.



\end{document}